# A Flood Routing Method for Data Networks

Jaihyung Cho

Monash University
Clayton 3168, Victoria
Australia
jaihyung@dgs.monash.edu.au

James Breen

Monash University
Clayton 3168, Victoria
Australia
jwb@dgs.monash.edu.au

## Abstract

In this paper, a new routing algorithm based on a flooding method is introduced. Flooding techniques have been used previously, e.g. for broadcasting the routing table in the ARPAnet [1] and other special purpose networks [3][4][5]. However, sending data using flooding can often saturate the network [2] and it is usually regarded as an inefficient broadcast mechanism. Our approach is to flood a very short packet to explore an optimal route without relying on a pre-established routing table, and an efficient flood control algorithm to reduce the signalling traffic overhead. This is an inherently robust mechanism in the face of a network configuration change, achieves automatic load sharing across alternative routes, and has potential to solve many contemporary routing problems. An earlier version of this mechanism was originally developed for virtual circuit establishment in the experimental Caroline ATM LAN [6][7] at Monash University.

## 1. Introduction

Flooding is a data broadcast technique which sends the duplicates of a packet to all neighboring nodes in a network. It is a very reliable method of data transmission because many copies of the original data are generated during the flooding phase, and the destination user can double check the correct reception of the original data. It is also a robust method because no matter how severely the network is damaged, flooding can guarantee at least one copy of the data will be transmitted to the destination, provided a path is available.

While the duplication of packets makes flooding a generally inappropriate method for data transmission, our approach is to take advantage of the simplicity and robustness of flooding for routing purposes. Very short packets are sent over all possible routes to search for the optimal route of the requested QoS and the data path is established via the selected route. Since the Flood Routing algorithm strictly controls the unnecessary packet duplication, the traffic overhead caused from the flooding traffic is minimal.

Use of flooding for routing purposes has been suggested before [3][4][5], and it has been noted that it can be guaranteed to form a shortest path route[10]. And an earlier protocol was proposed and implemented for the experimental local area ATM network (Caroline [6][7]). However the earlier protocol had problems with scaling timer values, and also required complex mechanism to solve potential race and deadlock problem. Our proposal greatly simplifies the previous mechanism and reduces the earlier problems.

Chapter 2 explains the procedure for route establishment and the simulation results are presented in chapter 3. The advantages of the Flood Routing are reviewed specifically in chapter 4. Chapter 5 concludes this paper with suggesting some possible application area and the future study issues.

## 2. Flood Routing Mechanism

Figure 1, 3, 4 show the stepwise procedure of the route establishment.

In the Figure 1, the host A is requesting a connection set up to the target host B. In the initial stage, a short connection request packet (CREQ) is delivered to the first hop router 1 and router 1 starts the flood of the CREQ packets.

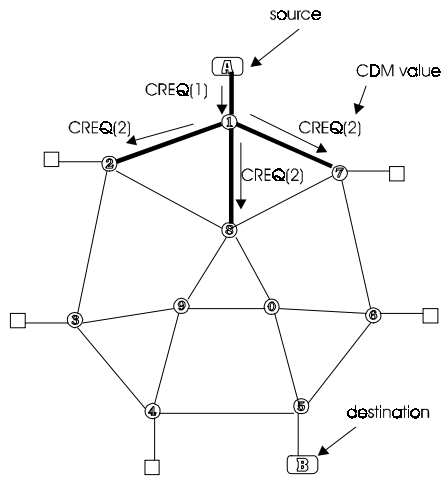

Figure 1

| |
|---|
| VC number (1byte=0) |
| Packet Type (1byte="CREQ") |
| CDM (1byte) |
| Source Address |
| Connection No (1byte) |
| Destination Address |
| QoS |

Figure 2   CREQ Packet Format

Figure 2 shows the format of the CREQ packet. The CREQ packet contains a connection difficulty metric (CDM) field, QoS parameters and the source & destination addresses and connection number. The metric can be any accumulative measure representing the route difficulty, such as hop count, delay, buffer length, etc. The connection number is chosen by the source host to distinguish the different packet floods of the same source and destination.

When a router receives the CREQ packet, the router matches the packet information with the internal Flood Queue to see if the same packet has been received before. If the CREQ packet is new, it records the information in the Flood Queue, increases the CDM value, and forwards the packet to all output links with adequate capacity to meet the QoS except the received one. Thus the flood of CREQ packets propagate through the entire network.

The Flood Queue is a FIFO list which contains the information relating to the best CREQ packet the router has received for each recent flood. As the flood packet of a new connection arrives and the information is pushed into the Flood Queue, the old information gradually moves to the rear and eventually is removed. The queueing delay from the insertion to the deletion depends on the queue size and the call frequency, and provided this delay is enough to cover the time for network wide flood propagation and reply, there is no need for a timer to wait to the completion of the flood.

Since the CDM value is increased as the CREQ packet passes the routers, the metric value represents the route difficulty that the CREQ packet has experienced. Because of the repeated duplication of the packet, a router may receive another copy of the CREQ packet. In this case, the router compares the metric values of the two packets and if the most recently arrived packet has the better metric value, it updates the information in the Flood Queue and repeats the flood action. Otherwise the packet is discarded. As a consequence, all the routers keep the record of the best partial route and the output link to use for setting up the virtual circuit.

Figure 3 shows the intermediate routers 2, 7, 8 have chosen the links toward the router 1 as the best candidate link. If one of them is requested for the path to the source node A, the router will use this link for the virtual circuit set up.

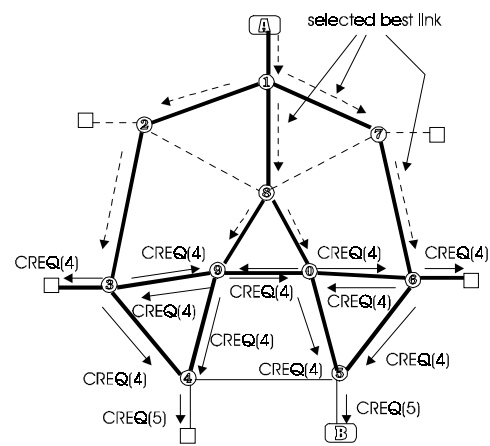

Figure 3

When the destination host receives a CREQ packet, it opens a short time-window to absorb possible further arriving CREQ packets. The expiration of the timer triggers the sending of the connection acceptance (CACC) packet along the best links indicated by the CREQ packet with the lowest CDM. The CACC packet is relayed back to the source host by the routers which at the same time install the virtual circuit via the optimal route. Finally, when the source host receives the CACC packet, the host may initiate data transmission.

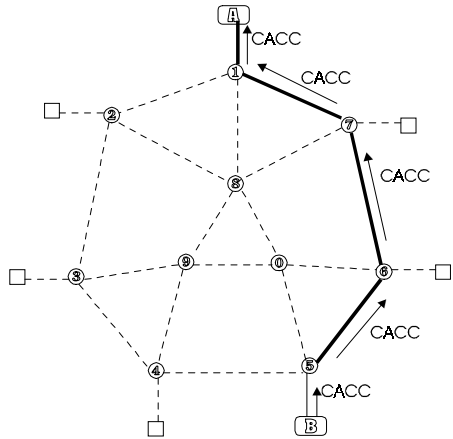

Figure 4

Note that bandwidth reservation occurs during the relay of the CACC packet. It is possible that the available QoS will have dropped below the requested level in one or more links. In this case, the source may either accept the lower QoS, or close the connection and try again.

More implementation details of the flooding protocol can be found in [9].

## 3. Simulation Result

One concern of Flood Routing is whether it will lead to congestion of the network by the signalling traffic. A simulation was carried out using various network conditions. Figure 5 shows the number of flooding packets produced in a connection trial in a normal traffic condition on a network consisting of 5 switching nodes, 9 hosts and 16 links. The simulation tested the event of 2000 seconds.

The graph shows that the total number of flooding packets per connection converges on the lower bound 18 with some exceptions. This is slightly higher than the number of the network links (16). This shows how the flood control mechanism is efficient in that the routers usually generate only one flooding packet per output link and this duplication process is rarely repeated again. As a result, the total number of flooding packets per connection is nearly same as the number of network links.

Considering the small size of the flooding packet, the bandwidth consumed by the signalling traffic is small. Suppose an ATM network using the Flood Routing generates 1000 calls per seconds, the bandwidth consumption by the signalling traffic will only be about 424 Kbps (= 1 K * 53 byte) per link and this does not include any additional route management traffic such as the routing table update.

From the simulation, it is observed that the average number and the maximum number of the flooding packets depends on the network topology and the traffic condition. If the network is simple topology such as a tree or a star shape, the average number of the flooding packets is nearly identical to the number of the network links. If the network is a complex topology such as a complete mesh topology, and there is a high traffic load, the routers tend to generate more packets because of the racing of the flooding packets.

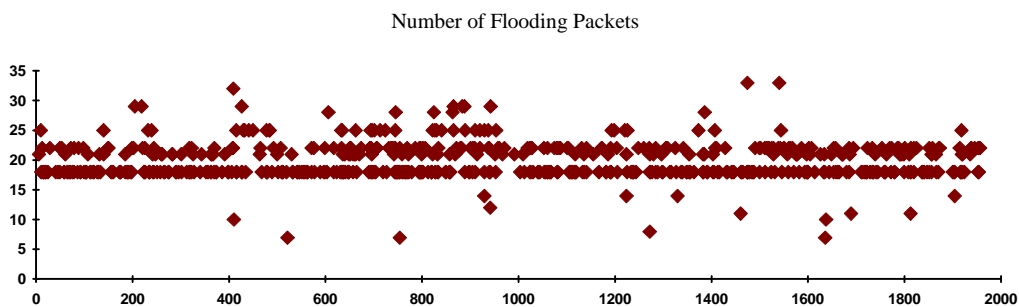

Number of Flooding Packets

Figure 5

The connections established by Flood Routing successfully avoid busy links and disperse the communication paths to all possible routes. This reduced the chance of congestion and utilizes all network resources efficiently.

## 4. Advantages of the Flood Routing

The distinctive features of the Flood Routing method are :

(a) It facilitates the load sharing of available network resources. If many possible routes exist between two end points in a network, the Flood Routing can disperse different connections over different routes to share the network load. Figure 6 shows this example.

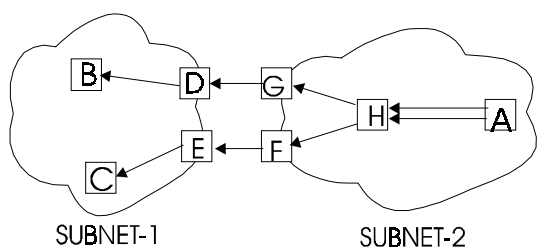

Figure 6   Example of Multipath Connection

In the sample network, there are more than two links exist between node A and H, and the node A used all links for different connections with balancing the load. More than two exterior routers are connecting the subnet 1 and the subnet 2, and the node H distributed the connections to all exterior routers. Therefore, all the network resources are utilized fully in Flood Routing network. This load sharing capability has been considered to be a difficult problem in table based routing algorithms.

(b) It automatically adapts to changes in the network configuration. For example, if the overall traffic between two end points has been increased, the network bandwidth can simply be expanded by adding more links between routers. The Flood Routing algorithm can recognize the additional links and use them for sharing the load in new connections.

(c) The method is robust. The Flood routing can achieve a successful connection even when the network is severely damaged, provided flooding packets can reach the destination. Once a flooding packet reaches the destination, the connection can be established via the un-damaged part of the network which was searched by the packet. This is very useful property in  networks which are vulnerable but which require high reliability, such as military networks.

(d) The method is simple to manage, as it makes no use of routing tables. This table-less routing method does not have the problem like "Convergence time" of the Distance Vector routing [8].

(e) It is possible to find the optimal route of the requested bandwidth or the quality of service. While the packet flood is progressing, bandwidth requirement and QoS constraints specified in the flooding packets are examined by the routers and the links that does not meet the requirements are excluded from the routing decision. As a result, the route constructed with the qualified links can meet the bandwidth and the QoS requirements, usually in the first attempt.

(f) It is a loop-free routing algorithm. The only possible case that the route may consist a loop can be caused from the corrupted metric information. However this can be detected by a check sum.

(g) Since the flooding method is basically a broadcast mechanism, it can be used for locating resources in network. Many network applications are best served by a broadcast facility, such as distributed data bases, address resolution, or mobile communications. Implementing broadcast in point-to-point networks is not straight forward. The flooding technique provides a means to solve this problem. In particular, locating a mobile user by Flood Routing, and establishing a dynamic route is an interesting issue. Application to a movable network in which entire network units including both the mobile users as well as the switching nodes and the wireless links is another potential research area.

## 5. Future Study and Conclusion

In this paper, we introduced a revised Flood Routing technique.  Flood Routing is a novel approach to network routing which has the potential to solve many of the routing problems in contemporary networks. The basic Flood Routing presented in this paper has been developed to be used in an ATM style network, however we believe a similar technique can also be applied to IP routing. Another promising area of application of this method would be military or mobile networks which require high mobility and reliability. Research to extend the point-to-point Flood Routing to optimal multi-point routing is now progressing. Further analysis of performance, and application to large scale networks are the future issues.